\documentclass[letterpaper]{article} 
\usepackage[preprint]{aaai2027}  
\usepackage[hyphens]{url}  
\usepackage{graphicx} 
\usepackage{natbib}  
\usepackage{caption} 
\usepackage{algorithm}

\usepackage{newfloat}
\usepackage{listings}
\DeclareCaptionStyle{ruled}{labelfont=normalfont,labelsep=colon,strut=off} 
\floatstyle{ruled}
\newfloat{listing}{tb}{lst}{}
\floatname{listing}{Listing}

\usepackage{booktabs}

\usepackage{array}       
\usepackage{algpseudocode}
\usepackage{amsmath}
\usepackage{amsthm}
\usepackage{amssymb}   
\usepackage{amsfonts}  
\usepackage{booktabs}
\usepackage{comment}
\usepackage{multirow}    
\usepackage{makecell}
\usepackage{soul}
\usepackage[switch]{lineno}
\usepackage{tabularx}
\usepackage[table]{xcolor} 
\usepackage[utf8]{inputenc}
\algrenewcommand\algorithmicrequire{\textbf{Input:}}
\algnewcommand\algorithmicparameter{\textbf{Parameter:}}
\algnewcommand\Parameter{\item[\algorithmicparameter]}
\algrenewcommand\algorithmicensure{\textbf{Output:}}
\algrenewcommand\algorithmiccomment[1]{\hfill{\footnotesize$\triangleright$~#1}}

\newcommand{\mymethod}[1]{\textbf{#1}} 

\newtheorem{theorem}{Theorem}
\newtheorem{definition}{Definition}

\title{FedGSA: Geometry-Consistent Subspace Aggregation for Differentially Private Federated LoRA}
\author{
    Lele Zheng,
    Ruijie Hu,
    Tao Zhang,
    Ke Cheng,
    Yulong Shen
}

\affiliations{
    Xidian University, China\\
}

\begin{document}

\maketitle

\begin{abstract}
Low-Rank Adaptation (LoRA) enables communication-efficient federated fine-tuning of pretrained language models. However, integrating differential privacy (DP) into federated LoRA remains challenging: independently perturbing and aggregating its two low-rank matrices can cause aggregation mismatch and the quadratic noise term. Existing methods mitigate these issues by freezing one low-rank matrix but still rely on Euclidean aggregation, which is basis-dependent and may distort the global update.
To address this limitation, we propose FedGSA, a geometry-consistent aggregation framework for differentially private federated LoRA. FedGSA represents each privatized client update as a basis-invariant subspace on the Grassmann manifold. In each communication round, clients extract low-dimensional subspaces capturing dominant update directions and encode them as projection matrices. The server aggregates these representations to estimate a geometry-consistent global update subspace and reconstructs the global LoRA factors within it, reducing distortion caused by basis misalignment, privacy noise, and heterogeneous client updates. We prove that FedGSA incurs no additional privacy loss beyond client-side DP training and establish its convergence under standard assumptions. Experiments on four GLUE tasks and a language generation benchmark demonstrate consistent improvements across privacy budgets and degrees of data heterogeneity. In particular, FedGSA improves average accuracy over the strongest baseline by 2.17\% and 2.27\% under $\epsilon=6$ and $\epsilon=3$, respectively.

\end{abstract}


\section{Introduction}

Pretrained language models often require task-specific fine-tuning to adapt to downstream tasks~\cite{yin2021comprehensive,zhang2023instruction}. When training data are distributed across clients, centralizing them may be infeasible due to privacy and regulatory constraints \cite{li2020federated}. Federated learning (FL) enables collaborative model adaptation without directly sharing local data, but full-parameter fine-tuning imposes substantial computation, storage, and communication costs on clients.
Parameter-Efficient Fine-Tuning (PEFT) methods~\cite{han2024parameter} offer a viable path by freezing most pre-trained parameters and optimizing only a small set of trainable ones, greatly reducing training and communication costs. Among these, LoRA enables efficient model adaptation by keeping the pre-trained model weights $W_0$ frozen and training only two low-rank matrices, $A \in \mathbb{R}^{r \times n}$ and $B \in \mathbb{R}^{m \times r}$. The update $\Delta W = BA$ dramatically reduces the number of trainable parameters, making it a representative solution for federated fine-tuning of large models.

Although federated LoRA keeps raw training data local, this alone does not eliminate privacy risks, as client updates may still reveal sensitive information about individual training records~\cite{zhu2019deep}. An honest-but-curious server may exploit these updates to infer record membership or sensitive attributes, or even reconstruct portions of the underlying training data~\cite{huang2021evaluating,geng2023improved}. Differential privacy (DP) provides a formal privacy guarantee by limiting the influence of each record on the training outcome~\cite{dwork2014algorithmic}. A widely adopted implementation is DP-SGD~\cite{abadi2016deep}, which clips per-example gradients to bound their sensitivity and injects calibrated Gaussian noise during local optimization. In federated LoRA,  clients can apply DP-SGD before transmitting their updates, thereby providing record-level protection against leakage through communicated model parameters. 

However, directly integrating DP-SGD with federated LoRA introduces challenges specific to its two-factor parameterization. First, directly averaging local LoRA factors is not equivalent to averaging the corresponding model updates, leading to aggregation mismatch and undesirable cross-client interaction terms, particularly under non-IID data distributions. Second, when DP noise is injected into both LoRA factors, the noise components interact through the matrix product, producing a quadratic term that can substantially amplify the effective noise and degrade model utility \citep{kang2024federated,sun2024improving}. Existing methods address aggregation mismatch and quadratic noise amplification through structural modifications to LoRA optimization and aggregation. FFA-LoRA~\cite{sun2024improving} addresses both issues by freezing one factor and optimizing only the other, but at the cost of reduced adaptation capacity. RoLoRA~\cite{chen2024robust}, LA-LoRA~\cite{liu2026rethinking}, and AS-LoRA~\cite{kim2026adaptive} alternately or adaptively update the factors; FedSVD~\cite{lee2025fedsvd}, FedASK~\cite{wen2025differentially}, and FLoRA~\cite{wang2024flora} instead employ SVD reparameterization, two-stage sketching, or matrix stacking to improve aggregation consistency. However, these designs either restrict adaptation capacity or introduce additional communication, computation, or estimation overhead. FedRot-LoRA~\cite{zhang2026fedrot} attributes the discrepancy between factor-wise averaging and correct update aggregation primarily to rotational ambiguity. Since, for any invertible matrix $Q$,
$
\Delta W = BA = (BQ)(Q^{-1}A).
$
Consequently, different clients may represent the same semantic update while embedding it in inconsistent latent subspaces. Under conditions of statistical heterogeneity, such uncertainty further interacts with client drift, thereby amplifying aggregation errors and degrading both training stability and global performance.

We further observe that the ambiguity of low-rank factorization arises from the coordinate-dependent treatment of LoRA factors in Euclidean space. To eliminate this basis dependence, it is necessary to move beyond factor-wise representations and adopt a basis-invariant subspace perspective, where equivalent low-rank updates are modeled as the same geometric object. Low-dimensional subspaces thus form a more fundamental aggregation target, with the Grassmann manifold providing their proper geometry and enabling consistent, invariant subspace-level aggregation.

Based on this insight, we propose FedGSA, a subspace aggregation method for differentially private federated LoRA. Instead of treating client LoRA updates as matrix representations in Euclidean space, FedGSA represents them as low-dimensional subspaces induced by their low-rank structure and aggregates them geometrically on the server. In each round, clients extract subspace representations reflecting principal update directions from their local LoRA updates. The server then aggregates these at the subspace level to estimate a globally consistent direction. The resulting consensus subspace further guides global-factor reconstruction, filtering inconsistent components before the reconstructed LoRA factors are broadcast to clients for subsequent local optimization. By operating at the subspace level, FedGSA avoids the basis dependency of direct averaging in coordinate space, making the aggregation invariant to basis transformations and more robust under differential privacy noise and data heterogeneity.

Our main contributions are summarized as follows:
\begin{itemize}
    \item We identify the representation dependence of factor-wise Euclidean aggregation in federated LoRA and introduce a geometric perspective that characterizes the dominant directions of client updates as basis-invariant subspaces on the Grassmann manifold.
    \item We propose FedGSA, a geometry-consistent aggregation framework for differentially private federated LoRA. FedGSA maps privatized client updates to projection representations, aggregates their dominant subspaces in a basis-invariant manner, and reconstructs the global LoRA factors within the resulting consensus subspace, reducing aggregation distortion under privacy noise and data heterogeneity.
    \item We theoretically show that FedGSA does not incur additional privacy loss and establish its convergence under standard assumptions. Experiments on four GLUE tasks and a language generation benchmark demonstrate consistent improvements across privacy budgets and degrees of data heterogeneity.
\end{itemize}

\section{Related Work}
\label{sec:relatedwork}

LoRA is particularly suitable for differentially private federated learning, as privacy mechanisms can be applied to compact low-rank adapters rather than the full model, substantially reducing computation and communication overhead. DP-LoRA~\cite{guo2024fedlfc,liu2025differentially} provides a straightforward solution by separately computing, clipping, and perturbing the per-sample gradients of the two factors, $A$ and $B$, before averaging the privatized updates at the server. However, this approach exposes two fundamental challenges~\cite{sun2024improving}. First, aggregating local weight updates is generally not equivalent to independently averaging their low-rank factors, resulting in an aggregation mismatch under naive factor-wise averaging~\cite{zhang2024towards}. Second, independently injected DP noise in $A$ and $B$ produces quadratic cross-noise terms when constructing the LoRA update $\Delta W=BA$, thereby amplifying the effective perturbation. 

FFA-LoRA~\cite{sun2024improving} mitigates both issues by freezing one factor, but limits adaptation capacity. To retain dual-factor expressiveness, FLoRA~\cite{wang2024flora} stacks client updates, FedSVD~\cite{lee2025fedsvd} privately updates $B$ and periodically reparameterizes the aggregated $BA$ via server-side SVD, and FedASK~\cite{wen2025differentially} approximates both factors through two-stage sketching, all at increased communication or computational cost. RoLoRA~\cite{chen2024robust} alternates factor optimization and aggregation across rounds, but suffers from stale-block errors and does not explicitly suppress DP noise. LA-LoRA~\citep{liu2026rethinking} instead alternates factors at each local step and low-pass filters privatized gradients, reducing DP perturbation and model sharpness. AS-LoRA~\citep{kim2026adaptive} adaptively selects $A$ or $B$ across layers and rounds using curvature, but incurs estimation overhead and overlooks inter-layer dependencies.

Despite these advances, DP-oriented methods address multiplicative cross-noise, factor coupling, or approximate reconstruction, but overlook basis ambiguity in privatized low-rank factors and cross-client subspace misalignment. FedRot-LoRA~\cite{zhang2026fedrot} identifies rotational invariance as causing aggregation error and aligns local factors to a global reference via orthogonal Procrustes transformations. However, it is limited to orthogonal reparameterizations, while the server still averages aligned $A$ and $B$ separately in Euclidean space; thus, general basis non-uniqueness and residual cross-client interactions remain. Moreover, FedRot-LoRA is not designed for differential privacy, where perturbations destabilize factor coordinates. These limitations motivate a basis-invariant geometric formulation. We represent each privatized LoRA update by its dominant subspace, aggregate these subspaces on the Grassmann manifold, and reconstruct global factors at the matrix level. This avoids coordinate-dependent factor averaging and mitigates factor misalignment, client heterogeneity, and privacy-induced distortion, while allowing both global factors to evolve across communication rounds.

\section{Preliminaries}
\label{sec:preliminaries}
The preceding discussion shows that the difficulty of differentially private federated LoRA arises not only from privacy noise, but also from applying coordinate-wise Euclidean aggregation to low-rank updates whose factorized representations are inherently non-unique. To formalize this mismatch and motivate a basis-invariant subspace perspective, we next review averaging in the ambient Euclidean matrix space and on the Grassmann manifold of $r$-dimensional subspaces.
\begin{definition}[Euclidean space]
A Euclidean space is typically a finite-dimensional real vector space $\mathbb{R}^n$ equipped with the standard inner product
$\langle x, y\rangle = x^\top y$, which induces the norm and distance$\|x\|_2=\sqrt{x^\top x}, d(x,y)=\|x-y\|_2$.
\end{definition}
For matrices, $\mathbb{R}^{m\times n}$ is algebraically and metrically equivalent to the standard Euclidean space $\mathbb{R}^{mn}$~\cite{horn2012matrix,golub2013matrix}. In such a geometry, the set of points forms a real vector space, and the operations of vector addition and scalar multiplication are well-defined. Meanwhile, the Euclidean inner product induces the canonical metric, under which the geodesic connecting two points is the straight line segment between them. Given a collection $\{X_i\}_{i=1}^N$, its arithmetic mean $\bar{X}=\frac{1}{N}\sum_{i=1}^N X_i$
is exactly the minimizer of the sum of squared Frobenius distances~\cite{boyd2004convex}:
\begin{equation}
\bar{X}=\arg\min_{Z}\sum_{i=1}^N \|X_i-Z\|_F^2.
\label{eq:Euclidean2}
\end{equation}

\begin{definition}[Grassmann manifold]
The Grassmann manifold (Grassmannian), denoted by $\mathrm{Gr}(k,n)$, is defined as the set of all $k$-dimensional linear subspaces of $\mathbb{R}^n$. This set admits a natural smooth manifold structure and can be equipped with a canonical Riemannian metric, thereby forming a compact Riemannian manifold.
\end{definition}
The Grassmannian $\mathrm{Gr}(k,n)$ admits multiple equivalent mathematical characterizations~\cite{edelman1998geometry,absil2008optimization}. One common quotient representation is
\begin{equation}
\mathrm{Gr}(k,n)\simeq \mathrm{St}(k,n)/\mathrm{O}(k),
\end{equation}
where $\mathrm{St}(k,n)$ denotes the \emph{Stiefel manifold}, i.e., the set of matrices $U\in\mathbb{R}^{n\times k}$ satisfying $U^\top U=I_k,$ and $\mathrm{O}(k)$ is the $k$-dimensional orthogonal group, corresponding to basis changes within the same subspace. Equivalently, $\mathrm{Gr}(k,n)$ can be expressed as a quotient of orthogonal groups:
\begin{equation}
\mathrm{Gr}(k,n)\simeq \mathrm{O}(n)/(\mathrm{O}(k)\times \mathrm{O}(n-k)).
\end{equation}

Another widely used representation is via projection matrices:
\begin{equation}
    \resizebox{.91\linewidth}{!}{$
            \mathrm{Gr}(k,n)\simeq \left\{ P\in\mathbb{R}^{n\times n}:\; P^2=P,\; P^\top=P,\; \mathrm{tr}(P)=k \right\},
        $}
\end{equation}%
namely the set of rank-$k$ symmetric orthogonal projectors.

In practical computations, a subspace $S\in \mathrm{Gr}(k,n)$ is typically represented by an orthonormal basis matrix $U\in\mathbb{R}^{n\times k}$ with $U^\top U=I_k$. Importantly, a point on the Grassmann manifold is not a specific basis $U$, but rather the subspace $\mathrm{span}(U)$. Therefore, for any orthogonal matrix $Q\in \mathrm{O}(k)$, the matrices $U$ and $UQ$ represent the same point on $\mathrm{Gr}(k,n)$.

\section{Method}
\label{sec:method}

\subsection{Design Rationale}

\begin{figure}[t]
  \centering
  \includegraphics[width=\columnwidth]{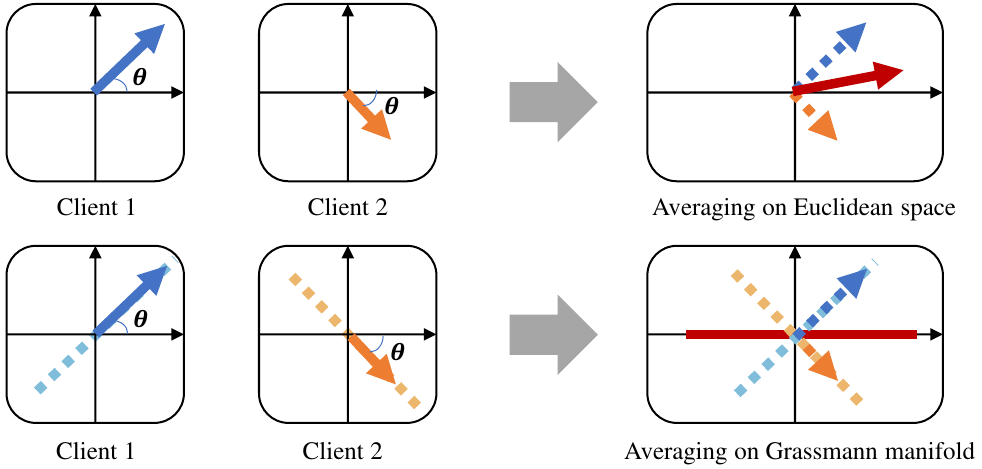}
  \caption{Illustration of Euclidean averaging versus Grassmann aggregation. Applying any nonzero scaling to a representative vector does not change the underlying subspace, but can substantially alter the direction of the Euclidean average.}
  \label{fig:motivation}
\end{figure}

FedGSA is grounded in the observation that the LoRA update $\Delta W=BA$ has an intrinsically low-rank structure, yet its factorized representation is non-unique. For any invertible matrix $Q$, we have
\begin{equation}
\Delta W=BA=(BQ)(Q^{-1}A).
\end{equation}
As a result, different clients may encode identical or similar update directions using different scales, signs, or bases. Direct Euclidean averaging therefore aggregates coordinate-dependent representations rather than the intrinsic low-rank structure, which may introduce basis-dependent bias. This issue can be illustrated by a one-dimensional subspace. In a two-dimensional plane, $u$ and $-u$ represent the same line passing through the origin, and hence the same subspace. However, their direct Euclidean average cancels out and yields the zero vector. Similarly, applying any nonzero scaling to a representative vector does not change the underlying subspace, but can substantially alter the direction of the Euclidean average as shown in Fig.~\ref{fig:motivation}. This shows that Euclidean averaging is highly sensitive to sign, scale, and basis choices, making it inadequate for stably characterizing the subspace consensus among client updates. This issue becomes more pronounced in federated settings, where client drift under non-IID data further interacts with stochastic perturbations introduced by DP-SGD. 

To remove this representation dependence, FedGSA treats each privatized LoRA update as a subspace-valued object. Specifically, the dominant column space of the local update matrix is regarded as an $r$-dimensional point on the Grassmann manifold, where different bases spanning the same subspace correspond to the same geometric object. Aggregating such subspaces enables the server to estimate a global directional consensus that is invariant to equivalent low-rank reparameterizations, rather than being sensitive to arbitrary factor coordinates.

\begin{figure*}[t]
  \centering
  \includegraphics[width=\textwidth]{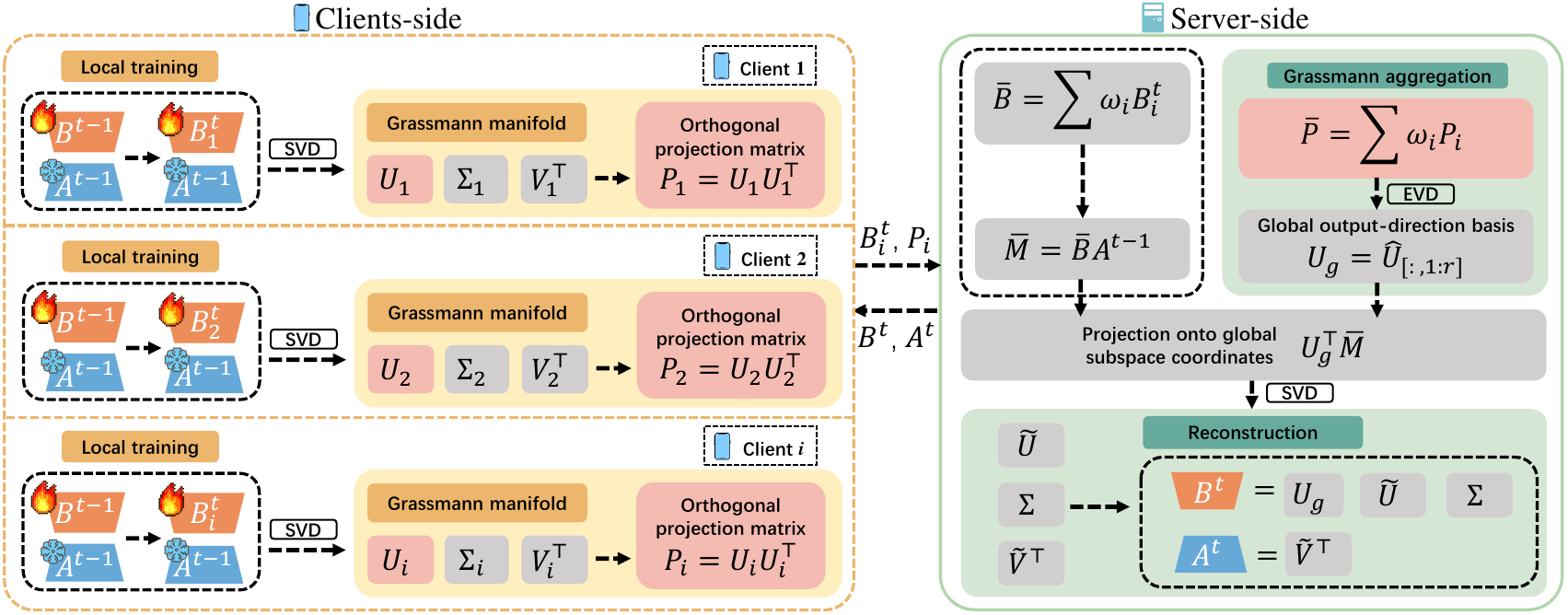}
  \caption{Overview of FedGSA. Each client maps its privatized LoRA update to a Grassmann point via orthogonal projection, while the server performs geometry-consistent subspace aggregation and reconstructs the global LoRA factors.}
  \label{fig:pipeline}
\end{figure*}

This rationale leads to a three-stage design. First, each client maps its privatized local LoRA update to a projection representation of its dominant update subspace. Second, the server performs geometry-consistent aggregation over these projection representations to estimate a global output-direction subspace. Third, the global LoRA factors are reconstructed within the aggregated subspace and broadcast to clients for the next communication round. Meanwhile, FedGSA mitigates aggregation bias and noise amplification by keeping $A$ fixed on each client and applying DP-SGD only to update $B$. Since subspace extraction, geometric aggregation, and low-rank reconstruction are all performed on already privatized client updates, they are post-processing operations and introduce no additional privacy loss. We outline our complete method in Fig.~\ref{fig:pipeline}.

\subsection{From LoRA Updates to Grassmann Points}

Given the representation ambiguity of low-rank factors, FedGSA does not directly aggregate the raw LoRA coordinates. Instead, it first converts each privatized local update into a basis-invariant subspace representation. Consider the $t$-th communication round. The server broadcasts the global LoRA factors from the previous round, where $A^{(t-1)}\in\mathbb{R}^{r\times d_{\mathrm{in}}}$ is fixed during local training. Each selected client $k$ locally optimizes $B_k^{(t)}\in\mathbb{R}^{d_{\mathrm{out}}\times r}$ using DP-SGD, where per-sample gradients are clipped and perturbed with Gaussian noise before being used for local updates.

After local training, client $k$ forms the privatized low-rank update matrix
\begin{equation}
    M_k^{(t)} = B_k^{(t)} A^{(t-1)}
    \in \mathbb{R}^{d_{\mathrm{out}}\times d_{\mathrm{in}}},
    \mathrm{rank}\big(M_k^{(t)}\big) \le r .
\end{equation}
While the entries of $M_k^{(t)}$ may be affected by local data heterogeneity and DP noise, its dominant column space captures the output-side update direction of client $k$. Therefore, rather than aggregating $M_k^{(t)}$ element-wise, FedGSA extracts its rank-$r$ subspace through a truncated singular value decomposition:
\begin{equation}
    M_k^{(t)} = U_k \Sigma_k V_k^\top,
    \qquad
    U_k \in \mathbb{R}^{d_{\mathrm{out}}\times r},
    \quad
    U_k^\top U_k = I .
\end{equation}
The corresponding orthogonal projection matrix is then defined as
\begin{equation}
    P_k = U_k U_k^\top .
\end{equation}
The matrix $P_k$ satisfies $P_k^\top=P_k$, $P_k^2=P_k$, and $\mathrm{rank}(P_k)=r$, and thus provides a canonical representation of an $r$-dimensional subspace in $\mathbb{R}^{d_{\mathrm{out}}}$. Since $P_k$ depends only on $\mathrm{span}(U_k)$ rather than the specific basis $U_k$, it is invariant to sign changes, rotations, and other equivalent basis transformations within the same subspace. Consequently, each privatized client update is represented as a point on the Grassmann manifold $\mathrm{Gr}(r,d_{\mathrm{out}})$, which serves as the geometric object to be aggregated by the server.

\subsection{Geometry-Consistent Subspace Aggregation}

Given the Grassmann representations $\{P_k\}_{k=1}^{K}$ of client updates, the server aims to estimate a global subspace that captures their shared update direction. Ideally, this can be formulated as a weighted Fréchet mean on the Grassmann manifold:
\begin{equation}
    P^\star
    =
    \arg\min_{P\in\mathcal{P}_r}
    \sum_{k=1}^{K} w_k d_G(P,P_k)^2 ,
\end{equation}
where $\mathcal{P}_r$ denotes the set of rank-$r$ orthogonal projection matrices in 
$\mathbb{R}^{d_{\mathrm{out}}\times d_{\mathrm{out}}}$, namely matrices satisfying 
$P^\top=P$, $P^2=P$, and $\mathrm{rank}(P)=r$; 
$d_G(\cdot,\cdot)$ denotes the geodesic distance on the Grassmann manifold, and 
$w_k\ge 0$ is the aggregation weight of client $k$.

Directly solving this intrinsic Fréchet mean generally requires iterative manifold optimization. To obtain an efficient and stable alternative, FedGSA adopts an extrinsic approximation based on the projection-matrix embedding. Specifically, the server first computes the weighted average of client projection matrices:
\begin{equation}
    \bar P = \sum_{k=1}^{K} w_k P_k .
\end{equation}
Although $\bar P$ is not necessarily a rank-$r$ projector, its dominant eigenspace provides a compact estimate of the global update subspace. We therefore perform eigenvalue decomposition (EVD)
\begin{equation}
    \bar P = U\Lambda U^\top,
    \qquad
    \Lambda=\mathrm{diag}(\lambda_1\ge\cdots\ge\lambda_{d_{\mathrm{out}}}),
\end{equation}
and take the leading $r$ eigenvectors as
\begin{equation}
    U_g = U_{[:,1:r]} .
\end{equation}

The resulting subspace $\mathrm{span}(U_g)$ serves as the geometry-consistent global output-direction subspace. Since the aggregation is performed over projection representations rather than raw factor coordinates, it depends only on the subspaces encoded by clients and is invariant to sign changes, rotations, and basis choices within each local update. This makes the global direction estimate more robust to representation mismatch, client heterogeneity, and stochastic perturbations introduced by differential privacy.

\subsection{Preconditioned Reconstruction of Global LoRA}

The aggregated Grassmann subspace provides a geometry-consistent estimate of the global output direction, but federated LoRA still requires explicit factors to be broadcast for the next communication round. FedGSA therefore reconstructs the global LoRA factors within the aggregated subspace, rather than directly factorizing a purely Euclidean average.

Given the global output subspace basis $U_g \in\mathbb{R}^{d_{\mathrm{out}}\times r}$, the server first forms the weighted matrix-level update
\begin{equation}
    \bar M^{(t)}
    =
    \sum_{k=1}^{K} w_k M_k^{(t)}
    =
    \sum_{k=1}^{K} w_k B_k^{(t)} A^{(t-1)} .
\end{equation}
Then, $\bar M^{(t)}$ is projected onto the aggregated output subspace:
\begin{equation}
    \widetilde M^{(t)}
    =
    U_g^\top \bar M^{(t)}
    \in \mathbb{R}^{r\times d_{\mathrm{in}}}.
\end{equation}
This projection acts as a subspace preconditioner: it filters the aggregated update along the geometry-consistent global direction, thereby suppressing components induced by heterogeneous drift, basis mismatch, or privacy noise.

We next perform a thin singular value decomposition on the reduced matrix:
\begin{equation}
    \widetilde M^{(t)}
    =
    \widetilde U \Sigma V^\top .
\end{equation}
The global LoRA factors are then updated as
\begin{equation}
    A^{(t)} = V^\top,
    \qquad
    B^{(t)} = U_g \widetilde U \Sigma .
\end{equation}
Consequently,
\begin{equation}
    B^{(t)}A^{(t)}
    =
    U_g \widetilde U \Sigma V^\top
\end{equation}
gives a rank-$r$ reconstruction of the aggregated update constrained to the global Grassmann subspace. This design preserves the intrinsic subspace consensus learned from client updates while producing standard LoRA factors for subsequent local optimization. Moreover, the orthogonal structure induced by $U_g$ and $V$ improves numerical conditioning and stabilizes the next round of federated training.

\begin{theorem}[Privacy Guarantee of FedGSA]
\label{thm:privacy_guarantee}
Assume that the $\ell_2$-sensitivity of the gradient computed in each local update is clipped to a constant $C = 1$.
In the $t$-th communication round, each sampled client performs DP-SGD during its local optimization, which is perturbed by Gaussian noise.
If the variance of the injected Gaussian noise $\sigma^2$ satisfies
\begin{equation}
\sigma^2=\mathcal{O}\!\bigl(
\tfrac{q_D^2 \cdot m \cdot q_K \cdot T \cdot \log(2/\delta)\cdot \log(2Tq_K/\delta)}
{\epsilon^2 \cdot K}
\bigr),
\end{equation}
where $q_K$ denotes the client sampling ratio in each communication round, $T$ is the total number of rounds, and $q_D$ is the data sampling ratio per local update, and where each client performs $m$ local updates per round, then over the joint client dataset $\mathcal{D}=\bigcup_{k=1}^K \mathcal{D}_k$, the final global LoRA factors $A^{(T)}$ and $B^{(T)}$ satisfy $(\epsilon,\delta)$-differential privacy.
\end{theorem} 

\begin{table*}[t]
    \centering
    \small
    \setlength{\tabcolsep}{3.5pt}
    \renewcommand{\arraystretch}{1.05}
    \begin{tabular}{@{}c|l|cccccc@{}}
        \toprule
        \multirow{2}{*}{\textbf{DP Budget}}
        & \multirow{2}{*}{\textbf{Method}}
        & \multicolumn{2}{c}{\textbf{MNLI}}
        & \multirow{2}{*}{\textbf{SST-2}}
        & \multirow{2}{*}{\textbf{QQP}}
        & \multirow{2}{*}{\textbf{QNLI}}
        & \multirow{2}{*}{\textbf{Average}} \\
        & & \textbf{Matched} & \textbf{Mismatched} & & & & \\
        \midrule

        \multirow{7}{*}{$\epsilon=6$}
        & FedAvg
        & 72.85 {\scriptsize$\pm$ 6.98}
        & 74.09 {\scriptsize$\pm$ 5.14}
        & 76.15 {\scriptsize$\pm$ 2.56}
        & 78.27 {\scriptsize$\pm$ 5.11}
        & 54.11 {\scriptsize$\pm$ 5.23}
        & 71.10 {\scriptsize$\pm$ 2.08}
        \\

        & FFA-LoRA
        & 64.30 {\scriptsize$\pm$ 8.63}
        & 66.55 {\scriptsize$\pm$ 8.91}
        & \underline{90.71} {\scriptsize$\pm$ 0.79}
        & 79.32 {\scriptsize$\pm$ 3.26}
        & 78.35 {\scriptsize$\pm$ 4.96}
        & 75.85 {\scriptsize$\pm$ 3.23}
        \\

        & FedSVD
        & \underline{75.20} {\scriptsize$\pm$ 2.31}
        & \underline{76.52} {\scriptsize$\pm$ 3.95}
        & 88.42 {\scriptsize$\pm$ 0.85}
        & \underline{80.02} {\scriptsize$\pm$ 2.51}
        & \underline{80.51} {\scriptsize$\pm$ 4.20}
        & \underline{80.13} {\scriptsize$\pm$ 1.21}
        \\

        & FedASK
        & 73.99 {\scriptsize$\pm$ 4.85}
        & 75.46 {\scriptsize$\pm$ 5.96}
        & 84.40 {\scriptsize$\pm$ 1.78}
        & 79.25 {\scriptsize$\pm$ 3.54}
        & 67.51 {\scriptsize$\pm$ 5.77}
        & 76.12 {\scriptsize$\pm$ 1.95}
        \\

        & LA-LoRA
        & 59.93 {\scriptsize$\pm$ 2.53}
        & 62.35 {\scriptsize$\pm$ 3.49}
        & 88.30 {\scriptsize$\pm$ 1.35}
        & 75.56 {\scriptsize$\pm$ 2.02}
        & 75.32 {\scriptsize$\pm$ 4.43}
        & 72.29 {\scriptsize$\pm$ 1.18}
        \\

        & AS-LoRA
        & 73.81 {\scriptsize$\pm$ 6.77}
        & 74.96 {\scriptsize$\pm$ 5.53}
        & 85.67 {\scriptsize$\pm$ 1.93}
        & 78.90 {\scriptsize$\pm$ 3.54}
        & 73.06 {\scriptsize$\pm$ 5.41}
        & 77.28 {\scriptsize$\pm$ 1.71}
        \\

        & \mymethod{FedGSA} (ours)
        & \textbf{77.36} {\scriptsize$\pm$ 3.07}
        & \textbf{77.96} {\scriptsize$\pm$ 4.03}
        & \textbf{91.98} {\scriptsize$\pm$ 0.15}
        & \textbf{82.13} {\scriptsize$\pm$ 2.76}
        & \textbf{82.06} {\scriptsize$\pm$ 3.21}
        & \textbf{82.30} {\scriptsize$\pm$ 2.25}
        \\

        \midrule

        \multirow{7}{*}{$\epsilon=3$}
        & FedAvg
        & 71.97 {\scriptsize$\pm$ 12.98}
        & 73.54 {\scriptsize$\pm$ 12.16}
        & 69.61 {\scriptsize$\pm$ 6.91}
        & 77.64 {\scriptsize$\pm$ 7.89}
        & 53.29 {\scriptsize$\pm$ 1.36}
        & 69.21 {\scriptsize$\pm$ 3.97}
        \\

        & FFA-LoRA
        & 50.73 {\scriptsize$\pm$ 7.62}
        & 52.25 {\scriptsize$\pm$ 7.09}
        & \underline{89.56} {\scriptsize$\pm$ 1.22}
        & 78.48 {\scriptsize$\pm$ 4.62}
        & 77.50 {\scriptsize$\pm$ 8.67}
        & 69.70 {\scriptsize$\pm$ 3.52}
        \\

        & FedSVD
        & 72.95 {\scriptsize$\pm$ 3.97}
        & 74.06 {\scriptsize$\pm$ 3.96}
        & 86.01 {\scriptsize$\pm$ 1.86}
        & 78.58 {\scriptsize$\pm$ 2.61}
        & \underline{79.75} {\scriptsize$\pm$ 1.87}
        & \underline{78.27} {\scriptsize$\pm$ 2.03}
        \\

        & FedASK
        & \underline{73.21} {\scriptsize$\pm$ 1.92}
        & \underline{74.26} {\scriptsize$\pm$ 4.38}
        & 76.95 {\scriptsize$\pm$ 0.55}
        & \underline{79.91} {\scriptsize$\pm$ 2.59}
        & 68.00 {\scriptsize$\pm$ 5.84}
        & 74.47 {\scriptsize$\pm$ 1.91}
        \\

        & LA-LoRA
        & 49.75 {\scriptsize$\pm$ 2.01}
        & 51.54 {\scriptsize$\pm$ 3.19}
        & 86.47 {\scriptsize$\pm$ 1.02}
        & 74.15 {\scriptsize$\pm$ 2.31}
        & 72.36 {\scriptsize$\pm$ 3.25}
        & 66.85 {\scriptsize$\pm$ 1.24}
        \\

        & AS-LoRA
        & 70.51 {\scriptsize$\pm$ 3.95}
        & 72.63 {\scriptsize$\pm$ 5.67}
        & 68.92 {\scriptsize$\pm$ 1.89}
        & 77.97 {\scriptsize$\pm$ 2.91}
        & 74.10 {\scriptsize$\pm$ 5.07}
        & 72.83 {\scriptsize$\pm$ 1.96}
        \\

        & \mymethod{FedGSA} (ours)
        & \textbf{74.54} {\scriptsize$\pm$ 2.73}
        & \textbf{75.22} {\scriptsize$\pm$ 3.05}
        & \textbf{90.93} {\scriptsize$\pm$ 1.02}
        & \textbf{80.98} {\scriptsize$\pm$ 4.45}
        & \textbf{81.03} {\scriptsize$\pm$ 3.99}
        & \textbf{80.54} {\scriptsize$\pm$ 2.54}
        \\

        \bottomrule
    \end{tabular}
    \caption{Results on five GLUE tasks with DP
    ($\epsilon\in\{3,6\}$, $\delta=10^{-5}$) under a non-IID Dirichlet
    split ($\alpha=0.5$). We report the average accuracy and 95\% confidence
    interval over three runs. The best and second-best results are highlighted
    in \textbf{bold} and \underline{underline}, respectively.}
    \label{tab:glue_private}
\end{table*}

\section{Experiments}

\subsection{Experimental Setups}
\paragraph{Datasets and Models.}
For natural language understanding tasks, we use the pretrained RoBERTa-base~\cite{liu2019roberta} model and conduct experiments on four datasets from the GLUE benchmark~\cite{wang2018glue}: MNLI~\cite{williams2018broad},  SST-2~\cite{socher2013recursive}, QQP~\cite{sharma2019natural} and QNLI. For generative tasks, We fine-tune GPT-2~\cite{radford2019language} on the E2E NLG Challenge dataset~\cite{novikova2017e2e}.

\paragraph{Baselines.}
We compare our proposed method FedGSA with the following six baselines:
\begin{enumerate}
    \item \textbf{FedAvg}~\cite{zhang2024towards,liu2025differentially}: Each client locally fine-tunes the LoRA matrices $A$ and $B$, and the server aggregates them via independent element-wise averaging.

    \item \textbf{FFA-LoRA}~\cite{sun2024improving}: The matrix $A$ is initialized using Kaiming Uniform in the range $(-d, d)$ and kept fixed throughout training; only the matrix $B$ is locally fine-tuned and aggregated.

    \item \textbf{FedSVD}~\cite{lee2025fedsvd}: Each client locally updates and uploads $B$. The server multiplies the aggregated $B$ with the previous-round $A$ to form $BA$, and then performs SVD-based reparameterization.

    \item \textbf{FedASK}~\cite{wen2025differentially}: FedASK proposes a two-stage sketching-based aggregation, achieving an accurate high-rank approximation of the aggregation target $\sum_k B_k A_k$ in a communication-efficient manner.

    \item \textbf{LA-LoRA}~\cite{liu2026rethinking}: Each client alternately updates $A$ and $B$ within each local round and applies low-pass smoothing to mitigate gradient coupling, DP noise amplification and aggregation instability.

    \item \textbf{AS-LoRA}~\cite{kim2026adaptive}: Each layer adaptively selects $A$ or $B$ across communication rounds using a curvature-aware score, without incurring additional privacy cost. 
\end{enumerate}

\subsection{Natural Language Understanding}
\paragraph{Experiments with DP-SGD.}
We evaluate all methods under $(\epsilon,\delta)$-DP with $\epsilon\in\{3,6\}$ and $\delta=10^{-5}$. As shown in Table~\ref{tab:glue_private}, FedGSA achieves the highest average accuracy under both privacy budgets, reaching $82.30\%$ at $\epsilon=6$ and $80.54\%$ at $\epsilon=3$, outperforming the strongest baseline, FedSVD, by $2.17\%$ and $2.27\%$, respectively. FedGSA also ranks first across all five evaluation tasks in both settings, demonstrating consistent robustness as the privacy constraint becomes stricter. We attribute these gains to single-factor private optimization, which avoids multiplicative cross-noise, and geometry-consistent Grassmann aggregation, which extracts basis-invariant directional consensus from noisy and heterogeneous client updates.
\paragraph{Experiments without privacy constraints.}
We further evaluate all methods on GLUE without differential privacy using RoBERTa-base under a non-IID Dirichlet split ($\alpha=0.5$). As shown in Table~2, FedGSA achieves the highest average accuracy of $88.81\%$, outperforming the strongest baseline, FedSVD, by $2.02\%$. It also ranks first on both MNLI splits and QNLI, while remaining competitive on QQP and SST-2. These results demonstrate that geometry-consistent subspace aggregation improves robustness to heterogeneous client updates even in the absence of privacy noise.

\begin{table*}[t]
    \centering
    \scriptsize
    \setlength{\tabcolsep}{3.0pt}
    \renewcommand{\arraystretch}{1.08}
    \resizebox{\textwidth}{!}{%
    \begin{tabular}{@{}l|cccccc|ccccc@{}}
        \toprule
        & \multicolumn{6}{c|}{\textbf{NLU: GLUE}}
        & \multicolumn{5}{c}{\textbf{NLG: E2E NLG Challenge}} \\
        \cmidrule(lr){2-7}
        \cmidrule(lr){8-12}

        \multirow{2}{*}[0.6ex]{\textbf{Method}}
        & \multicolumn{2}{c}{\textbf{MNLI}}
        & \multirow{2}{*}{\textbf{SST-2}}
        & \multirow{2}{*}{\textbf{QQP}}
        & \multirow{2}{*}{\textbf{QNLI}}
        & \multirow{2}{*}{\textbf{Average}}
        & \multirow{2}{*}{\textbf{BLEU} $\uparrow$}
        & \multirow{2}{*}{\textbf{NIST} $\uparrow$}
        & \multirow{2}{*}{\textbf{MET} $\uparrow$}
        & \multirow{2}{*}{\textbf{ROUGE-L} $\uparrow$}
        & \multirow{2}{*}{\textbf{CIDEr} $\uparrow$} \\
        & \textbf{Matched}
        & \textbf{Mismatched}
        & & & & & & & & & \\
        \midrule

        FedAvg
        & \underline{85.33} $\pm$ 13.97
        & 85.29 $\pm$ 15.22
        & \textbf{93.46} $\pm$ 13.12
        & \textbf{87.28} $\pm$ 10.32
        & 57.72 $\pm$ 12.51
        & 81.82 $\pm$ 7.18
        & 23.04
        & 1.87
        & 47.56
        & 49.81
        & 2.21
        \\

        FFA-LoRA
        & 82.07 $\pm$ \phantom{0}1.82
        & 82.50 $\pm$ \phantom{0}1.84
        & 93.69 $\pm$ \phantom{0}0.17
        & 84.78 $\pm$ \phantom{0}3.11
        & 89.69 $\pm$ \phantom{0}1.08
        & 86.55 $\pm$ 1.01
        & 27.51
        & 2.67
        & 52.21
        & 52.36
        & \underline{2.32}
        \\

        FedSVD
        & 82.73 $\pm$ \phantom{0}2.13
        & 82.79 $\pm$ \phantom{0}2.25
        & \underline{93.12} $\pm$ \phantom{0}0.53
        & 85.35 $\pm$ \phantom{0}2.35
        & \underline{89.97} $\pm$ \phantom{0}1.43
        & \underline{86.79} $\pm$ 1.62
        & 26.32
        & 2.18
        & 50.91
        & 52.58
        & 2.31
        \\

        FedASK
        & 85.25 $\pm$ \phantom{0}3.36
        & \underline{85.39} $\pm$ \phantom{0}2.91
        & 67.66 $\pm$ \phantom{0}1.09
        & 87.09 $\pm$ \phantom{0}2.66
        & 50.54 $\pm$ \phantom{0}2.64
        & 75.19 $\pm$ 2.17
        & \underline{29.61}
        & \underline{3.98}
        & \underline{54.05}
        & 50.01
        & 2.17
        \\

        LA-LoRA
        & 80.06 $\pm$ \phantom{0}2.21
        & 80.33 $\pm$ \phantom{0}1.52
        & 91.28 $\pm$ \phantom{0}0.56
        & 82.78 $\pm$ \phantom{0}2.54
        & 86.91 $\pm$ \phantom{0}1.40
        & 84.27 $\pm$ 1.66
        & 28.43
        & \textbf{4.46}
        & 53.67
        & 46.96
        & 1.83
        \\

        AS-LoRA
        & 81.34 $\pm$ \phantom{0}3.76
        & 82.00 $\pm$ \phantom{0}3.01
        & 90.48 $\pm$ \phantom{0}0.99
        & 84.44 $\pm$ \phantom{0}2.40
        & 88.17 $\pm$ \phantom{0}2.75
        & 85.29 $\pm$ 2.23
        & 27.69
        & 2.60
        & 52.34
        & \underline{52.79}
        & \underline{2.32}
        \\

        \midrule
        \textbf{FedGSA} (ours)
        & \textbf{87.01} $\pm$ \phantom{0}2.95
        & \textbf{87.57} $\pm$ \phantom{0}2.52
        & 92.01 $\pm$ \phantom{0}0.21
        & \underline{87.12} $\pm$ \phantom{0}3.59
        & \textbf{90.33} $\pm$ \phantom{0}1.73
        & \textbf{88.81} $\pm$ 1.85
        & \textbf{30.32}
        & 3.08
        & \textbf{54.56}
        & \textbf{53.45}
        & \textbf{2.40}
        \\

        \bottomrule
    \end{tabular}%
    }
    \caption{NLU results on five GLUE tasks without privacy constraints under
    a non-IID Dirichlet split ($\alpha=0.5$). NLG results are obtained using
    GPT-2 on the E2E NLG Challenge with DP
    ($\epsilon\in\{3,6\}$, $\delta=10^{-5}$) under the same data split.}
    \label{tab:nlu_nlg_results}
\end{table*}

\begin{figure}[t]
  \centering
  \includegraphics[width=\columnwidth]{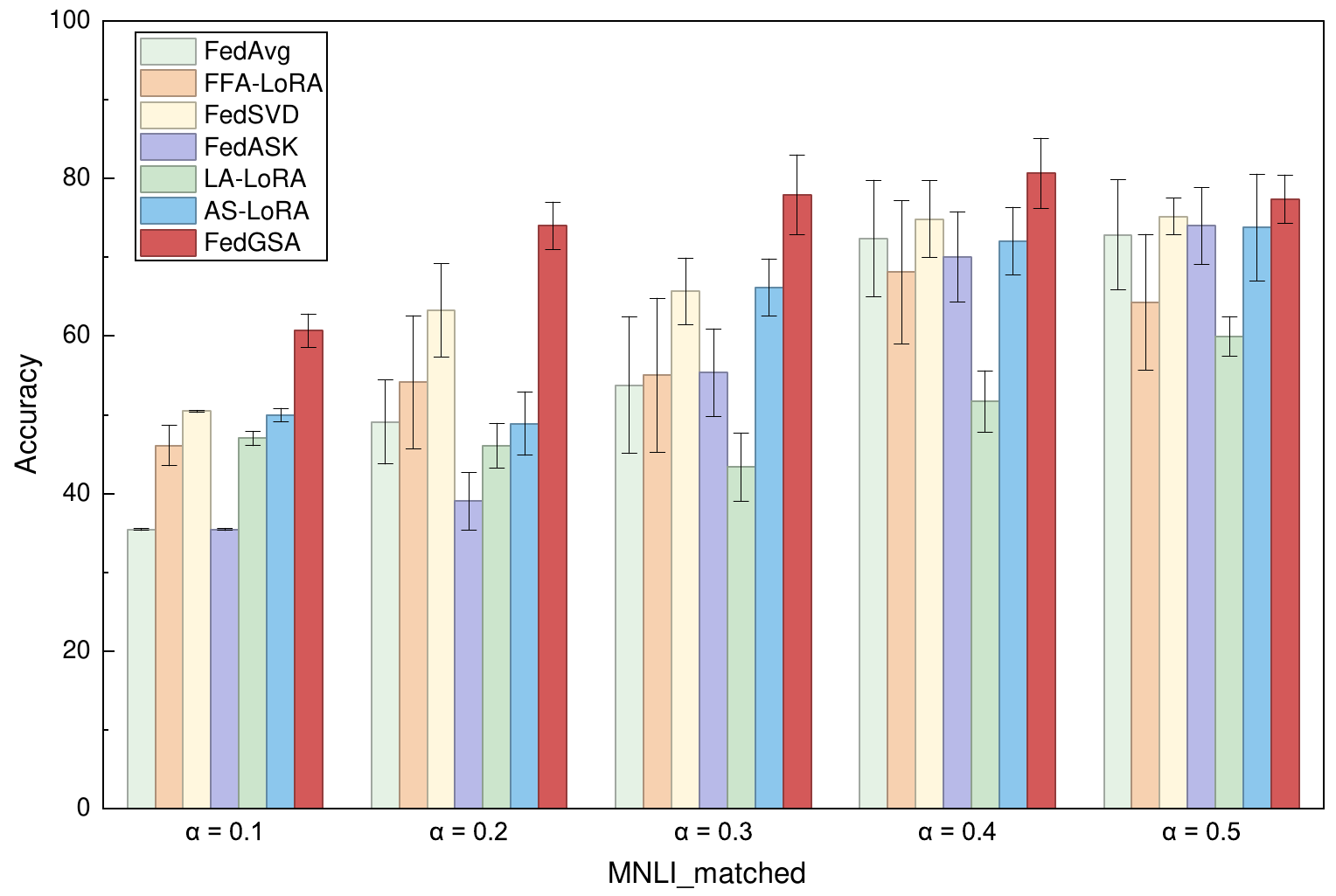}
  \caption{Results of varying $\alpha \in \{ 0.1,0.2,0.3,0.4,0.5 \}$ for a Dirichlet distribution on the MNLI dataset.}
  \label{fig:dirichlet}
\end{figure}

\paragraph{Heterogeneity of the Data Distribution ($\alpha$).}
We evaluate robustness to statistical heterogeneity on MNLI using Dirichlet splits with $\alpha \in \{0.1,0.2,0.3,0.4,0.5\}$, where smaller $\alpha$ indicates stronger non-IID skew. All methods are trained with DP-SGD under $(\epsilon,\delta)=(6,10^{-5})$, and the mean accuracy with 95\% confidence intervals over five runs is reported. As shown in Fig.~\ref{fig:dirichlet}, FedGSA achieves the highest accuracy across all settings, outperforming the strongest baseline by $2.16\%$--$11.82\%$. Its advantage is particularly pronounced at smaller $\alpha$, demonstrating that geometry-consistent subspace aggregation better preserves shared update directions under severe client heterogeneity.

\subsection{Natural Language Generation}
We evaluate FedGSA on GPT-2 using the E2E NLG Challenge under the private federated setting. As shown in Table~\ref{tab:nlu_nlg_results}, FedGSA achieves the best performance on four of five metrics at rank $r=4$, reaching 30.32 BLEU, 54.56 METEOR, 53.45 ROUGE-L, and 2.40 CIDEr, while remaining competitive on NIST. The consistent gains across lexical-overlap and semantic-adequacy metrics indicate that subspace aggregation retains informative generation directions despite privacy perturbations and heterogeneous client updates during training. These results demonstrate that FedGSA generalizes beyond discriminative NLU tasks and preserves strong generation utility under differential privacy.

\subsection{Ablation Study}

\paragraph{Effect of Grassmann Aggregation.}
\begin{figure}[t]
  \centering
  \includegraphics[width=\columnwidth]{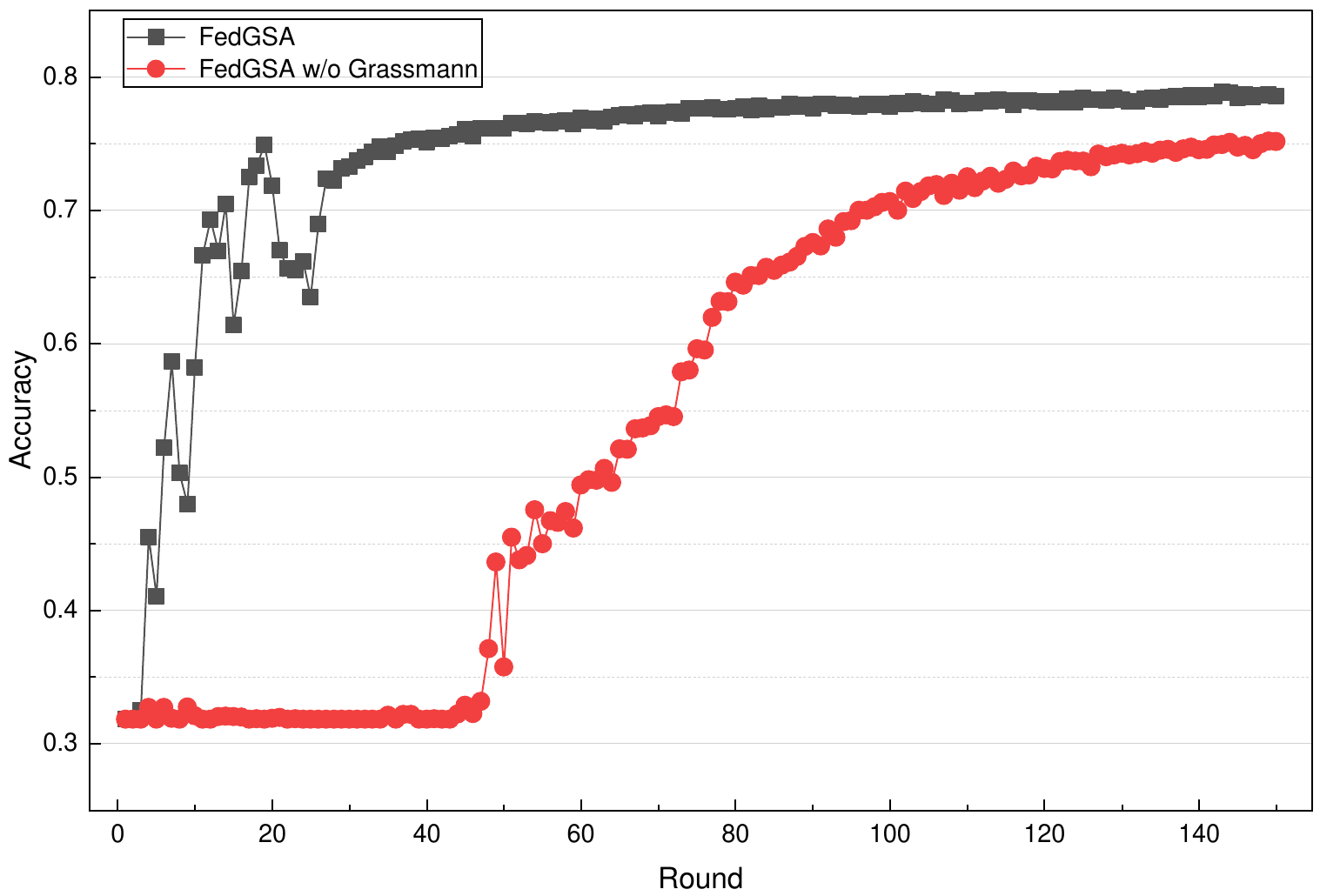}
  \caption{Ablation study of Grassmann aggregation on MNLI with RoBERTa-base under DP ($\epsilon = 6$, $\delta=10^{-5}$) and a non-IID Dirichlet split ($\alpha=0.5$).}
  \label{fig:ablation_grassmann}
\end{figure}
We construct FedGSA w/o Grassmann, which retains single-factor DP-SGD but replaces projection-based subspace aggregation with truncated SVD of the Euclidean-averaged update. All other settings remain unchanged. As shown in Fig.~\ref{fig:ablation_grassmann}, FedGSA reaches $70\%$ accuracy by round 14, whereas the ablated variant requires 96 rounds. At round 150, FedGSA achieves $78.61\%$, outperforming the variant by $3.43\%$; over the final 20 rounds, their average accuracies are $78.55\%$ and $74.68\%$, respectively. These results demonstrate that Grassmann aggregation better preserves basis-invariant directional consensus under privacy noise and client heterogeneity, thereby improving convergence and final utility.

\paragraph{Interaction between Local Optimization and Aggregation.}
\begin{table}[t]
\centering
\small
\setlength{\tabcolsep}{5pt}
\renewcommand{\arraystretch}{1.15}

\begin{tabular}{lcc}
\toprule
\textbf{}
& \textbf{Euclidean Aggregation}
& \textbf{Grassmann Aggregation} \\
\midrule
$A$ and $B$
& 72.85
& 73.97 \\
Freeze $A$
& 64.30
& \textbf{77.36} \\
\bottomrule
\end{tabular}
\caption{Ablation of client optimization and server aggregation on MNLI-matched under DP ($\epsilon=6$, $\delta=10^{-5}$) and a non-IID Dirichlet split ($\alpha=0.5$).}
\label{tab:optimization_aggregation_ablation}
\end{table}
We conduct a $2\times2$ ablation study to examine the interaction between client-side LoRA optimization and server-side aggregation. As shown in Table~\ref{tab:optimization_aggregation_ablation}, Grassmann aggregation improves accuracy by $1.12\%$ when both $A$ and $B$ are updated, but yields a substantially larger gain of $13.06$ points when $A$ is frozen. Moreover, freezing $A$ decreases accuracy by $8.55$ points under Euclidean aggregation, whereas it improves accuracy by $3.39$ points under Grassmann aggregation. The complete FedGSA configuration therefore achieves the best accuracy of $77.36\%$. These results reveal a strong interaction between the two components: single-factor private optimization suppresses multiplicative DP noise, while geometry-consistent aggregation preserves directional consensus and alleviates the expressiveness loss induced by fixing one LoRA factor.

\section{Conclusion}
In this work, we studied differentially private federated LoRA from a basis-invariant geometric perspective. We identified that factor-wise Euclidean aggregation is sensitive to the non-uniqueness of low-rank representations, causing cross-client subspace misalignment under privacy noise and data heterogeneity. To address this issue, we proposed FedGSA, which combines single-factor private optimization with geometry-consistent aggregation on the Grassmann manifold. FedGSA maps privatized client updates to projection representations, estimates a consensus update subspace, and reconstructs global LoRA factors within that subspace. We proved that these server-side operations introduce no additional privacy loss and established convergence under standard assumptions. Experiments on GLUE and E2E NLG demonstrate consistent gains across privacy budgets and heterogeneity levels, while ablations confirm the complementary roles of noise suppression and subspace aggregation for robust federated adaptation.

\clearpage
\bibliography{aaai2027}

\end{document}